\documentstyle[12pt] {article}
\begin {document}
\parindent=15pt
\begin{center}
\vskip 1.5 truecm
{\bf HEAVY QUARK PAIR CORRELATIONS IN QCD}\\
\vspace{.5cm}
Yu.M.Shabelski and A.G.Shuvaev \\
\vspace{.5cm}
Petersburg Nuclear Physics Institute, \\
Gatchina, St.Petersburg 188350 Russia \\
\end{center}
\vspace{1cm}
\begin{abstract}
The azimutal correlations of heavy quarks produced in the high energy $pp$
($p\overline{p}$) collisions are calculated in the framework of QCD
without the usual assumptions of the parton model. The virtual nature of the
interacting gluons as well as their transverse motion and different
polarizations are taken into account. We give some predictions for the
azimutal correlations of charm and beauty hadrons produced at
Tevatron-collider and LHC.
\end{abstract}
\vspace{3cm}

E-mail SHABEL@VXDESY.DESY.DE \\

E-mail SHUVAEV@THD.PNPI.SPB.RU

\newpage

\section{INTRODUCTION}
The investigation of the heavy quarks production in the high energy
hadron processes provides a method for studying the internal structure
of hadrons. Realistic estimates of the cross section of the heavy quark
production as well as their correlations are necessary in order to plan
experiments on existing and future accelerators. These predictions are
obtained usually in the parton model framework and depend significantly
on the quark and gluon structure functions. The last ones are more or
less known experimentally from the data of HERA, but unknown at the very
small values of Bjorken variable $x < 10^{-4}$. However it is just the
region that dominates in the heavy quark production at the high energies.

The Gribov-Lipatov-Dokshitzer-Altarelli-Parisi (GLDAP) evolution
equation is applied usually to calculate the structure functions. It
sums up in the leading logarithm approximation (LLA) all the QCD diagram
contributions proportional to $(\alpha_s\ln\,q^2)^n$, but it does not
take into account the terms proportional to $(\alpha_s\ln\,1/x)^n$.
That is why this approximation does not give the correct asymptotic
behaviour of the
structure function in the small $x$ region. For the correct description
of these phenomena not only the terms of the form $(\alpha_s\ln\,q^2)^n$
have to be collected in the Feynman diagrams but also the terms
$(\alpha_s\ln\,1/x)^n$.

Another problem that appears at $x \sim 0$ is that of the absorption
(screening) corrections which must stop the increase of the cross
section at $x\rightarrow 0$ in accordance with the unitarity
condition. It can be interpreted as the saturation of parton density.
For relatively small virtuality $q^2\leq\,q_0^2 (x)$ the gluon
structure function behaves as $xG(x,q^2)\sim q^2R^2$, so the cross
section for the interaction of the point-like parton with a target,
$\sigma\sim(1/q^2)xG(x,q^2)\sim R^2$, obeys the unitarity condition. The
quantity $q_0^2(x)$ can be treated as a new infrared-cutoff parameter
which plays the role of the typical transverse momentum of partons in the
parton cascade of the hadron in semihard processes. The behaviour of
$q_0(x)$ was discussed in ref.[6]. It increases with $\log(1/x)$ and
the values of $q_0(x)$ are about 2-4 GeV at $x = 0.01-0.001$.

The predictions for the cross section of the heavy quark pair
production are based usually on the parton model calculations [1, 2].
All particles are assumed in this model to be on mass shell
with the longitudinal component of the momentum only (so called collinear
approximation), and the cross section is averaged over two transverse
polarizations of the gluons. The virtualities $q^2$ of the initial
partons are taken into account only through their densities. The latter
are calculated in LLA through GLDAP evolution equation. The
probabilistic picture of noninteracting partons underlies this way of
proceeding. In the region, where the transverse mass $m_T$ of the
produced heavy quark is close to $q_0(x)$, the dependence of the
amplitude of the subprocess $gg \rightarrow \overline{Q} Q$,
dominating at the high energy, on the virtualities and polarizations
of the gluons becomes important. The amplitudes of
these subprocesses should be calculated more accurately than
in the parton model. The matrix elements of the QCD subprocesses
accounting for the virtualities and polarizations of the gluons are very
complicated. We presented them in our previous paper [4]
for the main and simplest subprocesses, $gg \rightarrow \overline{Q} Q
\; (\sim \alpha_{s}^{2})$ for hadroproduction and $\gamma g
\rightarrow \overline{Q} Q \; (\sim \alpha_{s})$ for photo- and
electroproduction. The contributions of high-order subprocesses
can be essential, but our aim is to discuss the qualitative
difference between our results and the parton model predictions.
This can be done on the level of the low-order diagrams.

In this paper we calculate in the same order ($\sim \alpha_{s}^{2}$) the
azimutal correlations of the heavy quarks produced in hadron-hadron
collisions which are in serious disagreement with the conventional parton
model predictions (see \cite{BEAT} and references therein).

\section{CROSS SECTIONS OF HEAVY FLA\-VO\-UR PRODUCTION IN QCD}

The cross section of heavy quark hadroproduction is given schematically
by the diagrams in Fig.1. The main contribution to the cross
section at small $x$ is known to come from the gluons. The lower and
upper ladder blocks represent the two-dimensional gluon
distributions $\varphi(x,q_1^2)$ and  $\varphi(x,q_2^2)$, which
are the functions of the longitudinal momentum fraction
($x$ and $y$) of the initial hadron and the virtuality of the gluon.
Their distribution over $x$ and transverse momenta
$q_T$ in hadron is given in semihard theory \cite{GLR} by the
function $\varphi(x,q^2)$. It differs from the usual function
$G(x,q^2)$:  \begin{equation} \label{xg} xG(x,q^2) \,=\,
\frac{1}{4\sqrt{2}\,\pi^3} \int^{q^2}_0 \varphi
(x,q^2_1)\,dq_1^2. \end{equation} Such a definition of
$\varphi(x,q^2)$ enables to treat correctly the effects
arising from the gluon virtualities. The exact expression for this
function can be obtained as a solution of the evolution equation
which, contrary to the parton model case, is nonlinear due to
interactions between the partons in the small $x$ region.

In what follows we use Sudakov decomposition for the quark momenta
$p_{1,2}$ through the momenta of the colliding hadrons  $p_A$ and
$p_B\,\, (p^2_A = p^2_B \simeq 0)$  and the transverse ones $p_{1,2T}$:
\begin{equation}
\label{1}
p_{1,2} = x_{1,2} p_B + y_{1,2} p_A + p_{1,2T}.
\end{equation}
The differential cross sections of heavy quarks hadroproduction have the
form:\footnote{We put the argument of $\alpha_S$ to be equal to the gluon
virtuality, which is very close to the BLM scheme\cite{blm}; (see also
\cite{lrs}).}
$$ \frac{d\sigma_{pp}}{dy^*_1 dy^*_2 d^2 p_{1T}d^2
p_{2T}}\,=\,\frac{1}{(2\pi)^8}
\frac{1}{(s)^2}\int\,d^2 q_{1T} d^2 q_{2T} \delta (q_{1T} +
q_{2T} - p_{1T} - p_{2T}) $$
\begin{equation}
\label{spp}
\times\,\,\frac{\alpha_s(q^2_1)}{q_1^2} \frac{\alpha_s (q^2_2)}{q^2_2}
\varphi(q^2_1,y)\varphi (q^2_2, x)\vert M_{QQ}\vert^2.
\end{equation}
Here $s = 2p_A p_B\,\,$, $q_{1,2T}$ are the gluon transverse momenta
and $y^*_{1,2}$  are the quark rapidities in the hadron-hadron c.m.s.
frame,
\begin{equation}
\label{xy}
\begin{array}{crl}
x_1=\,\frac{m_{1T}}{\sqrt{s}}\, e^{-y^*_1}, &
x_2=\,\frac{m_{2T}}{\sqrt{s}}\, e^{-y^*_2},  &  x=x_1 + x_2\\
y_1=\, \frac{m_{1T}}{\sqrt{s}}\, e^{y^*_1}, &  y_2 =
\frac{m_{2T}}{\sqrt{s}}\, e^{y^*_2},  &  y=y_1 + y_2. \end{array}
\end{equation}
$\vert M_{QQ}\vert^2$ is the square of the matrix element for
the heavy quark pair hadroproduction.

In LLA kinematics
\begin{equation}
\label{q1q2}
\begin{array}{crl}
q_1 \simeq \,yp_A + q_{1T}, & q_2 \simeq \,xp_B + q_{2T},
\end{array}
\end{equation}
so
\begin{equation}
\label{qt}
\begin{array}{crl}
q_1^2 \simeq \,- q_{1T}^2, & q_2^2 \simeq \,- q_{2T}^2.
\end{array}
\end{equation}
(The more accurate relations are $q_1^2 =- \frac{q_{1T}^2}{1-y}$,
$q_2^2 =- \frac{q_{2T}^2}{1-x}$, but we are working in the kinematics,
where $x,y \sim 0$).

The matrix element $M_{QQ}$ is calculated in the Born order of
QCD without standart simplifications of the parton model, since
in the small $x$ domain there are no grounds for neglecting the
transverse momenta of the gluons $q_{1T}$ and $q_{2T}$ in
comparision with the quark mass and the parameter $q_0(x)$. In
the axial gauge $p^\mu_B A_\mu = 0$ the gluon propagator takes
the form $D_{\mu\nu} (q) = d_{\mu\nu} (q)/q^2,$ \begin{equation}
\label{prop}
d_{\mu\nu}(q)\, =\, \delta_{\mu\nu} -\, (q^\mu p^\nu_B \,+\, q^\nu p^\mu_B
)/(p_B q).
\end{equation}
For the gluons in $t-$channel the main contribution comes from the so
called 'nonsense' polarization  $g^n_{\mu\nu}$, which can be picked out
by decomposing the numerator into the longitudinal and the transverse parts:
\begin{equation}
\label{trans}
\delta_{\mu\nu} (q)\, =\, 2(p^\mu_B p^\nu_A +\, p^\mu_A p^\nu_B
)/s\, +\, \delta^T_{\mu\nu} \approx\, 2p^\mu_B p^\nu_A
/s\,\equiv\, g^n_{\mu\nu}.  \end{equation} The other
contributions are suppressed by the powers of $s$. Since the sum
of the diagrams in Fig.1a-1c is gauge invariant in the LLA, the
transversality condition for the ends of gluon line enables one to
replace  $p^\mu_A$  by  $-q^\mu_{1T}/x$  in the expression for
$g^n_{\mu\nu}$. Thus we get
\begin{equation}
\label{trans1}
d_{\mu\nu} (q)\,\, \approx\,\, -\,2\, \frac{p^\mu_B q^\nu_T}{x\,s},
\end{equation}
or
\begin{equation}
\label{trans2}
d_{\mu\nu} (q)\,\, \approx\,\, \,2\, \frac{q^\mu_T q^\nu_T}{xys},
\end{equation}
if we do such a trick for the vector $p_B$ too. Both these equations for
$d_{\mu\nu}$  can be used, but for the form (\ref{trans1}) one has to
modify slightly the gluon vertex (to account for the several ways of gluon
emission --- see ref. \cite{3} ):
\begin{equation}
\label{geff}
\Gamma_{eff}^{\nu} =
\frac{2}{xys}\,[(xys - q_{1T}^2)\,q_{1T}^{\nu} - q_{1T}^2
q_{2T}^{\nu} + 2x\,(q_{1T}q_{2T})\,p_B^{\nu}].
\end{equation}
As a result the colliding gluons can be treated as aligned ones and
their polarization vectors are directed along the transverse momenta.
Ultimately, the nontrivial azimuthal correlations must arise between
the transverse momenta $p_{1T}$ and $p_{2T}$ of the heavy quarks.

Formally, one looses the gauge invariance dealing with off
mass shell gluons ($q_1,q_2$). Indeed, the new graphs similar to
the 'bremsstruhlung' from the initial (final) quark line (Fig.1d)
may contribute to the central plato rapidity region in the covariant
Feynman gauge. However this is not the fact. The
function $\phi(x,q^2)$ collects with the "semihard" accuracy all
terms of the form $\alpha_s^k(\ln q^2)^n(\ln (1/x))^m$
with $n+m\ge k$.
In this case the triple gluon vertex (11) includes effectively all the
main (leading logarithmic) contributions of the type of Fig.1d
\cite{BFKL}. For example, the upper part of the graph shown in Fig.1d
corresponds in terms of the BFKL equation to the t-channel gluon reggeization.
Therefore the final expression is gauge invariant
with the logarithmic accuracy.

Although the situation considered here seems to be quite opposite to the
parton model, there is a certain limit in which our formulae can be
transformed into the parton model ones. Consider the $pp$ case and
assume that the characteristic values of quark momenta $p_{1T}$
and $p_{2T}$ are much larger than the values of gluon momenta,
$q_{1T},q_{2T}$,
\begin{equation}
\label{par1}
<p_{1T}> \gg <q_{1T}> \;, \;\; <p_{2T}> \gg <q_{2T}>,
\end{equation}
and one can keep only the lowest powers of $q_{1T}, q_{2T}$. It means that
we can put $q_{1T} = q_{2T} = 0$ everywhere in the matrix element $M_{QQ}$
except the vertices. Introducing the polar coordinates
\begin{equation}
\label{pol}
d^2q_{1T}\, =\, \frac{1}{2}\, dq^2_{1T} d\theta_1
\end{equation}
(and the same for $q_{2T}$) and performing angular integration with the
help of the formula
\begin{equation}
\label{aint}
\int^{2\pi}_0 d\theta_1 q^\mu_{1T} q^\nu_{1T} \, =\,\, \pi\, q^2_{1T}
\delta^{\mu\nu}_T
\end{equation}
we obtain
\begin{equation}
\label{part}
\int^{2\pi}_0\, d\theta_1 \frac{q^\mu_{1T}}{y}\frac{q^\nu_{1T}}{y}\,
\int^{2\pi}_0\, d\theta_2 \frac{q^\lambda_{2T}}{x}
\frac{q^\sigma_{2T}}{x} M_{\mu\nu}\overline{M}_{\lambda\sigma}\,
=\,2\pi^2 \frac{q^2_{1T}
q^2_{2T}}{(x\,y)^2}\,\vert M_{part}\vert^2.
\end{equation}
Here $M_{part}$ is just the parton model matrix element, since the
result is the same as that calculated for the real (mass shell) gluons
and averaged over transverse polarizations. Then we obtain the cross
section (\ref{spp}) in the form [2, 3]:
$$ \frac{d\,\sigma}{dy_1^* dy^*_2 d^2 p_{1T}} = $$ $$=\vert
M_{part}\vert^2 \frac{1}{(\hat{s})^2} \int \frac{\alpha_s (q^2_1)
\varphi(y,q^2_{1T})}{4\,\sqrt{2}\, \pi^3}
\frac{\alpha_s(q^2_2)\varphi (x,q^2_{2T})}{4
\,\sqrt{2}\, \pi^3}\, dq^2_{1T}\,dq^2_{2T}\,=$$
\begin{equation}
\label{pcs}
=\,\frac{\alpha_s^2 (q^2)}{(\hat{s})^2}\,\vert M_{part}\vert^2\,
xG(x,q^2_2)\,yG(y,q^2_1),
\end{equation}
where $\hat{s}=xys$  is the mass square of $\overline{Q} Q$ pair.

On the other hand, the assumption (\ref{par1}) is not fulfilled in
a more or less realistic case. The transverse momenta of produced quarks
as well as the gluon virtualities (QCD scale values) should be of the order
of heavy quark masses.

\section{RESULTS OF CALCULATIONS}

Eq.(3) enables to calculate straightforwardly the distributions over
the azimutal angle $\phi$, which is defined as the opening angle
between the two produced heavy quarks projected onto a plane
perpendicular to the beam.  Let us define this plane as
$xy$-plane and let the first heavy quark flights to the $x$
direction, i.e. $p_{1Tx} = p_{1T}$, $p_{1Ty} = 0$.  In this case
$\cos{\phi} = p_{2Tx}/p_{2T}$, and it is not a problem to
evaluate the distribution over $\phi$ integrating
Eq.(3) using, say, VEGAS code.

Since the functions $\varphi (x,q^2_2)$ and $\varphi (y,q^2_1)$ are
unknown at the small values of $q^2_2$ and $q^2_1$, we rewrite the
integrals in Eq.(3) as
$$ \int^{\infty}_{0} \, d q^2_{iT}
\delta (q_{iT} - p_{1T} - p_{2T})\, \delta(y_1 + y_2 - 1) \frac{\alpha_s
(q^2_i)}{q^2_i} \varphi (q^2_i, x)\vert M \vert^2 \frac{1}{y_2} = $$
\begin{equation}
\label{int}
= 4\sqrt{2}\,\pi^3 \delta (p_{1T} + p_{2T})\, \delta(y_1 + y_2 - 1)
\alpha_s (Q^2_0) xG(x,Q^2_0)\,\frac{1}{y_2}\,
(\frac{\vert M \vert^2}{q^2_i})_{q_i\rightarrow 0}
\end{equation}
$$\,+\, \int^{\infty}_{Q^2_0} \, d q^2_{iT}
\delta (q_{iT} - p_{1T} - p_{2T})\, \delta(y_1 + y_2 - 1)
\frac{\alpha_s (q^2_i)}{q^2_i}
\varphi (q^2_i, x)\vert M \vert^2 \frac{1}{y_2},  $$
where Eq.(\ref{xg}) is used\footnote{Here the value $Q^2_0$ should not
be mixed with the function $q^2_0(x)$ discussed in the
Introduction.}. Thus, we obtain the sum of the three different contributions:
the product of two first terms with $i = 1, 2$ in the r.h.s. of Eq.(17),
$w_1(\phi)$; the sum of the products of first and second terms,
$w_2(\phi)$; and, finally, the product of second terms, $w_3(\phi)$. The
first contribution, $w_1(\phi)$, is very similar to the conventional LO
parton model expression, in which the total heavy quark
momenta is exactly zero and the angle between quarks is always $180^o$.
However the angle between two heavy hadrons can be sligtly different
from this value due to a hadronization process. To take it into account
we assume that in the first contribution, where quarks are produced
back-to-back, the probability to find the final azimutal angle
$\phi$ ,
\begin{equation}
\sin \phi = \frac{p_h}{\sqrt{p_h^2 + p_T^2}} \;,
\end{equation}
is determined by the expression
\begin{equation}
w (p_h) = \frac{2}{\sqrt{\pi}p_0} e^{-p_h^2/p_0^2}  \;,
\end{equation}
where $p_0$ = 0.2 GeV/c and $p_h$ is the transverse momentum in the azimutal
plane of the hadron coming from hadronization process. The two
last contributions, $w_2(\phi)$ and $w_3(\phi)$, result in a
more or less broad distribution over the angle between the
produced quarks, so we neglect here their small modification in the
hadronization.

It is necessary to repeat that our approach is justified only at small
$x$, that is at the high enough initial energies. Unfortunately,
the highest energy, where the experimental data on the charm azimutal
correlations exist, is only $\sqrt{s}$ = 39 GeV \cite{BEAT}. The values
of $q_1^2$ and $q_2^2$ in Eq.(3), giving an essential contributions to
the charm production cross section, are not large enough at this energy,
so the first contribution in Eq.(17) dominates at the not very small
$Q^2_0$ value and our predictions coincide practically with the results
of LO parton model.

At the higher energies the second and third contributions become large
enough and lead to some difference with the conventional parton
model. The predictions of charm pair azimutal correlations at
$\sqrt{s}$ = 1800 GeV and $Q^2_0$ = 4 GeV$^2$ are presented in Fig.2a.
The solid hystogram shows the results for all produced charm particles.
However in this case some hadronization mechanism can contribute to the
azimutal correlations of the charmed hadrons with the small relative momenta.
To decrease this possible contribution the dashed histogram presents the same
results for the events, where both charmed hadrons have the transverse
momenta $p_T > p_{T min}$, $p_{T min}$ = 4 GeV/c.

The same results at the energy $\sqrt{s}$ = 14 TeV are shown in Fig.2b.
Note that the predicted azimutal correlations have some energy
dependence, they become more broad, when the initial energy increases,
however this energy dependence is weak enough, especially at
$p_{T min}$ = 0. Our $\phi$-distributions seem to be more broad
than the results of parton model calculations.

The similar predictions for the beauty pair azimutal correlations at the
energies $\sqrt{s}$ = 1800 GeV and 14 TeV are presented in Fig.3 for the
two values of $p_{T min}$, 0 and 8 GeV/c. For both these $p_{T min}$ values
the distributions become more broad with the energy increase.

Contrary to the parton model results, we predict two peaks for all
cases, the standard one at $\phi = 180^o$ and the second peak with the
smaller altitude, at $\phi = 0^o$, originated from the contribution of
the diagram Fig.1c.

The HERA experimental data on $F_2(x,Q^2)$ at the small x, actually
$10^{-4} < x < 10^{-2}$, show the singular $x$-behaviour at the
moderate $Q^2$ (say, $Q^2 \sim 10^1$ GeV$^2$). Both ZEUS
\cite{ZEUS} as well as H1 \cite{H1} Coll. data can be
parametrized as $x^{-\delta}$ with $\delta = 0.1 \div 0.25$.
Such a behaviour at $x \to 0$ is in evident contradiction with
the unitarity and has to be stopped by a shadow mechanism
\cite{GLR,MQ,KMSR}.

To see the possible influence of the shadow effects on the azimutal
correlations we make the simplest assumption that the shadowing modifies
the gluon distribution in such a way that the real distribution can be
written as
\begin{equation}
xG(x,Q^2) = \frac{xG_0(x,Q^2)}{1 + \epsilon xG_0(x,Q^2)} \;,
\end{equation}
where $\epsilon \ll 1$ and $xG_0(x,Q^2)$ is the bare GRV (HO) gluon
distribution \cite{GRV}. Both  $xG(x,Q^2)$ and $xG_0(x,Q^2)$
distributions are shown in Fig.4 for the value $\epsilon = 0.01$. One
can see that the difference between them at the smallest $x$, where the
data exist ($x \sim 10^{-4}$), is about 10 \%.

The calculated results for the azimutal correlations of the heavy flavour
pairs with "shadowed" gluon distributions, Eq.(20), are presented in Fig.5
and show that the shadowing does not significantly affect our results.

\section{CONCLUSION}

The above discussion shows that the accounting for the virtual nature of
the interacting gluons as well as their transverse motion and different
polarizations results in the qualitative discrepances with the parton model
predictions. Using, say, LO QCD formulae we obtain a considerably more
broad distribution over the angle between two heavy flavours than the
convenient NLO parton model (see \cite{BEAT}). The reason is that the
NLO parton model allows the discussed distributions to differ from
$\delta$-functions only due to the emission of one hard gluon.
Our approach, to the contrary, incorporates effectively the
emission of all evolution gluons via the phenomenological gluon
distribution.

We are grateful to M.G.Ryskin for useful discussions and to
E.M.Levin who participated at the early stage of the
calculations.

This work is supported in part by INTAS-93-0079, Russian Fund of
Fundamental Research (95-2-03145) and the Volkswagen Stifung.

\newpage

\begin{center}
{\bf Figure captions}\\
\end{center}

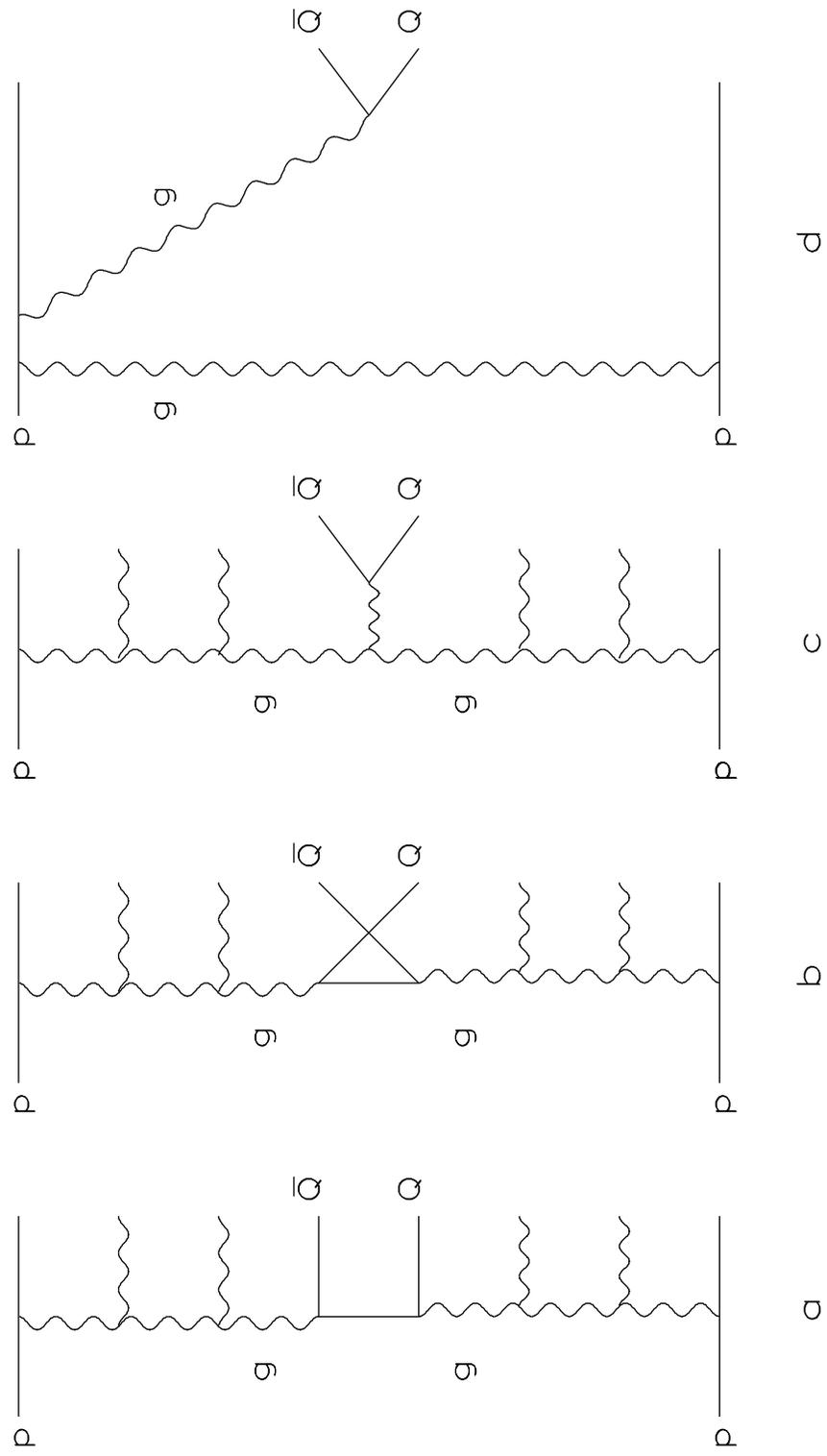
Fig. 1. Low order QCD diagrams for the heavy quark production in $pp$
($p\overline{p}$) collisions via gluon-gluon fusion (a-c) and the
diagram (d) violating formally the gauge invariance, which is
restored, however, with logarithmic accuracy.

Fig. 2. The calculated azimutal correlations of the charm pair production
in $pp$ ($\overline{p}p$) collisions at $\sqrt{s}$ = 1800 GeV (a) and
$\sqrt{s}$ = 14 TeV (b) for all events (solid histograms) and for the events,
where both the charm particles have $p_T > p_{T min}$ (dashed hystograms).

Fig. 3. The calculated azimutal correlations of the beauty pair production
in $pp$ ($\overline{p}p$) collisions at $\sqrt{s}$ = 1800 GeV (a) and
14 TeV (b) for all events (solid histograms) and for the events, where both
the beauty particles have $p_T > p_{T min}$ (dashed histograms).

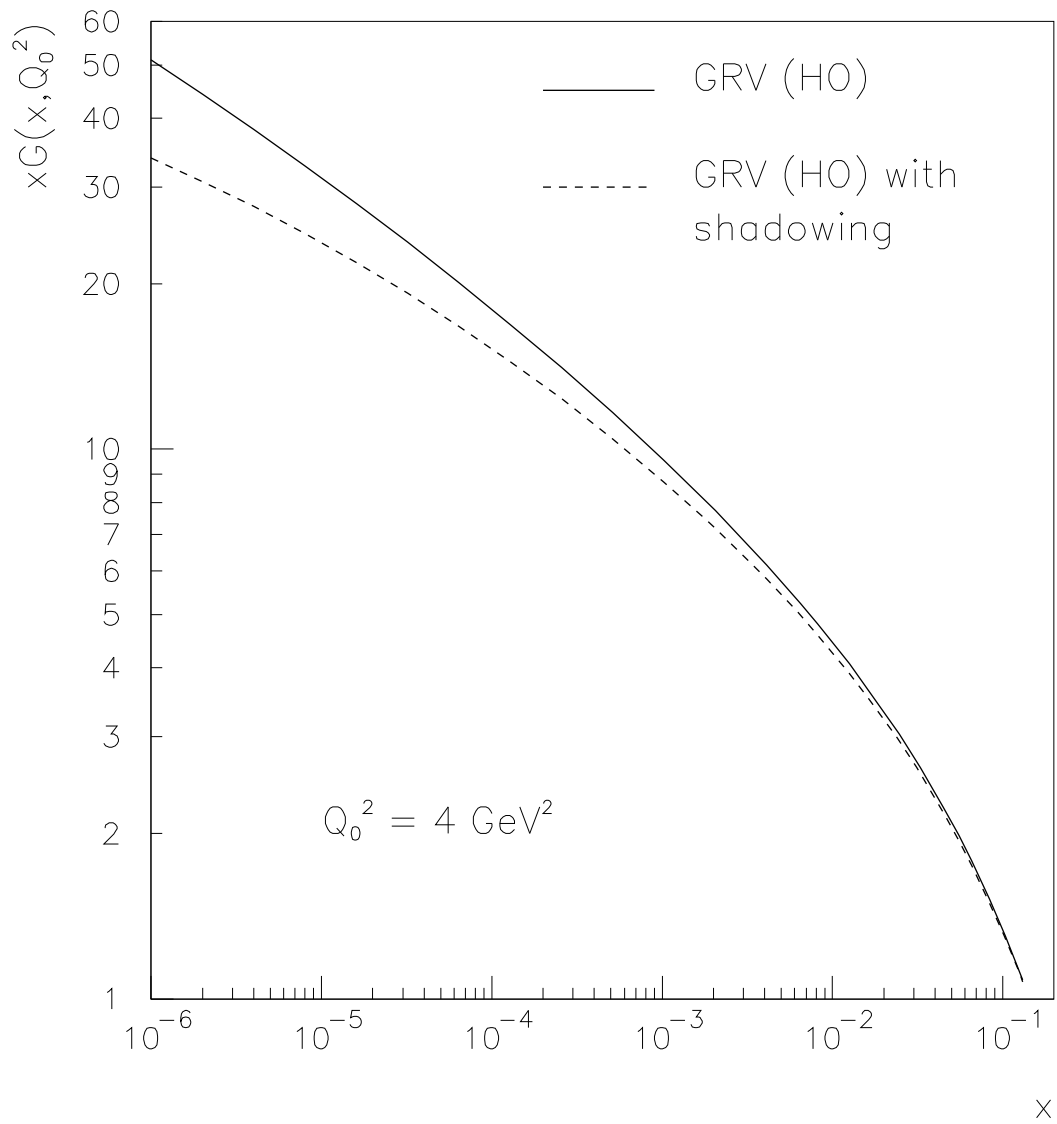
Fig. 4. Gluon structure function of the nucleon \cite{GRV} without (solid
curve) and with (dashed curve) shadow correction, Eq.(19).

Fig. 5. The calculated azimutal correlations of charm (a) and beauty (b)
pair production in $pp$ ($\overline{p}p$) collisions at $\sqrt{s}$ = 14
TeV for all events (solid histograms) and for the events, where both
the heavy flavour particles have $p_T > p_{T min}$ (dashed histograms).


\newpage

\begin{thebibliography}{99}
\bibitem{GLR} L.V.Gribov, E.M.Levin and M.G.Ryskin. Phys.Rep. 100 (1983)
1; E.M.Levin and M.G.Ryskin. Phys.Rep. 189 (1990) 267.
\bibitem{1} P.Nason, S.Dawson and R.K.Ellis. Nucl.Phys. B303 (1988) 607.
\bibitem{2} G.Altarelli. Nucl.Phys. B308 (1988) 724.
\bibitem{3} M.G.Ryskin, Yu.M.Shabelski and A.G.Shuvaev. Z.Phys. C69 (1996)
269.
\bibitem{BEAT} BEAT Coll., M.Adamovich et al. Phys. Lett. B348 (1995)
256.
\bibitem{blm} S.J.Brodsky, G.P.Lepage and P.B.Mackenzie. Phys.Rev. D28
(1983) 228.
\bibitem{lrs} E.M.Levin, M.G.Ryskin, Yu.M.Shabelski and A.G.Shuvaev.
Yad.Fiz. 54 (1991) 1420.
\bibitem{3} E.M.Levin, M.G.Ryskin, Yu.M.Shabelski and A.G.Shuvaev.
Yad.Fiz. 53 (1991) 1059.
\bibitem{BFKL} E.A.Kuraev, L.N.Lipatov, V.S.Fadin: Sov. Phys. JETP 45
(1977) 199.
\bibitem{GRV} Gluck, Reya and A.Vogt. Z.Phys. B263 (1991) 579.
\bibitem{ZEUS} ZEUS Coll., M.Derrick et al. Phys. Lett. B345 (1995) 576.
\bibitem{H1} H1 Coll., S.Aid et al. Phys. Lett. B354 (1995) 494.
\bibitem{MQ} A.H.Mueller and J.Qiu. Nucl.Phys. B268, (1986) 427.
\bibitem{KMSR} J.Kwiechinski, A.D.Martin, W.J.Stirling and R.G.Roberts.
Phys.Rev. D42 (1990) 3645.
\end{thebibliography}
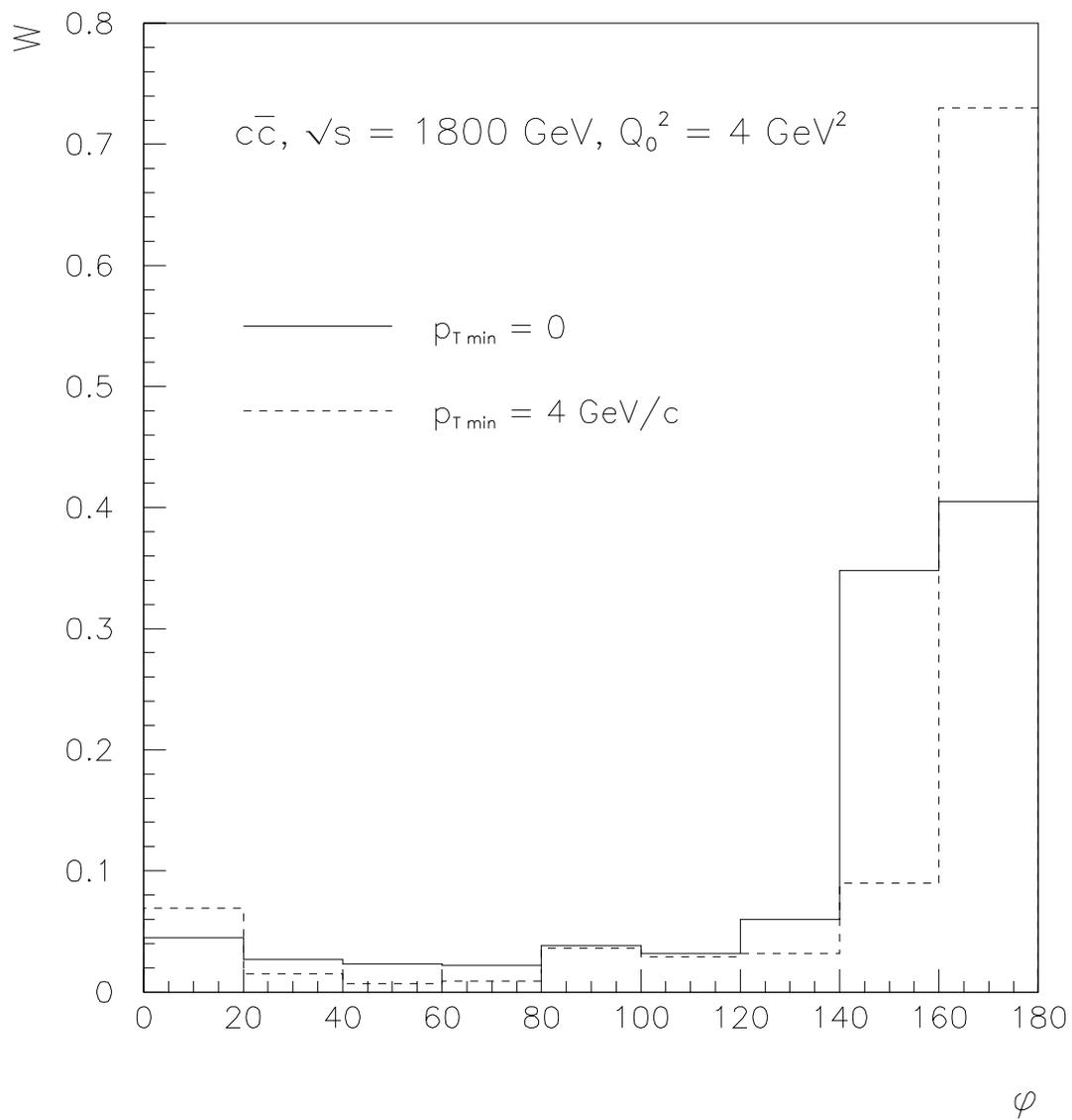
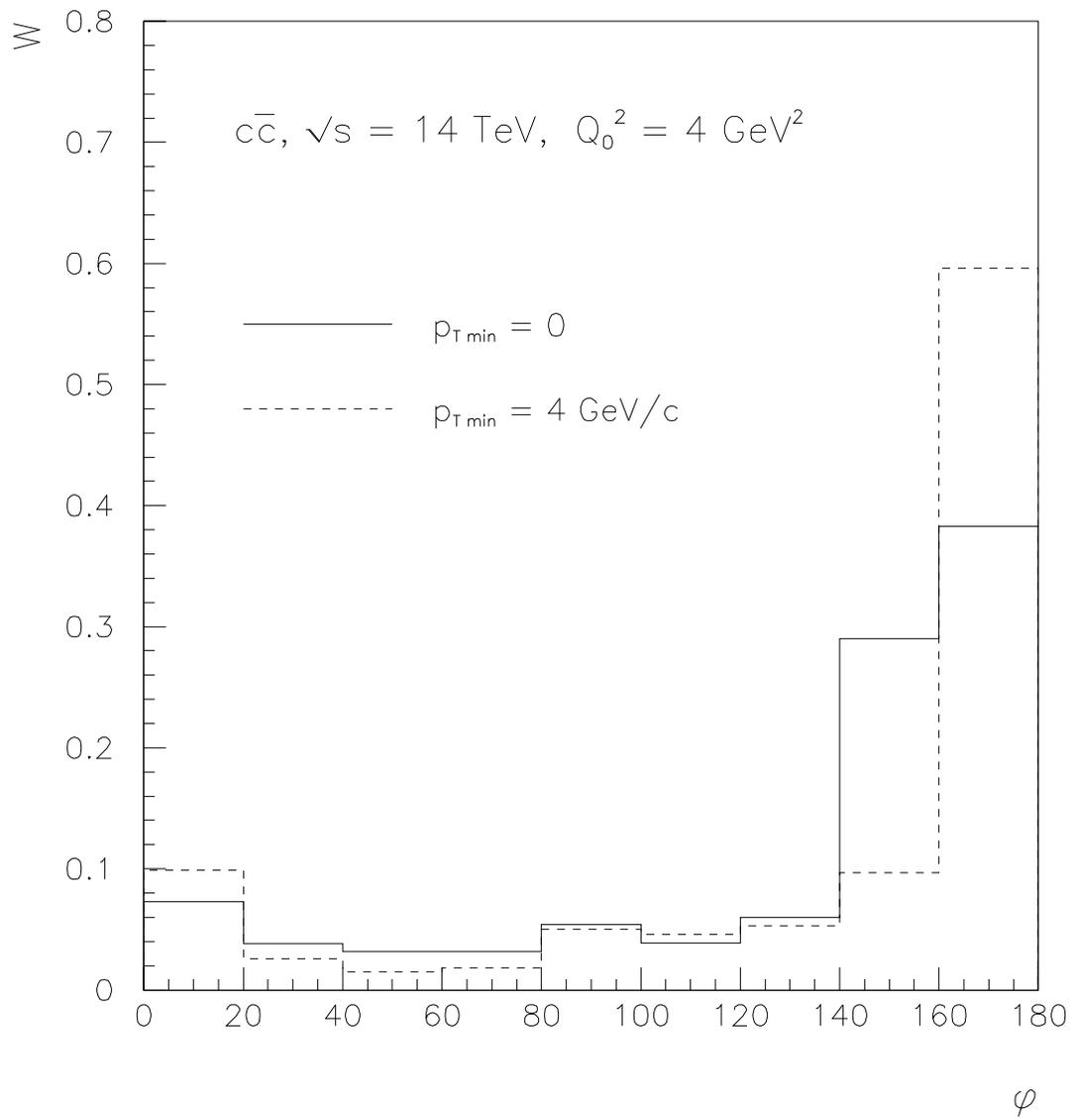
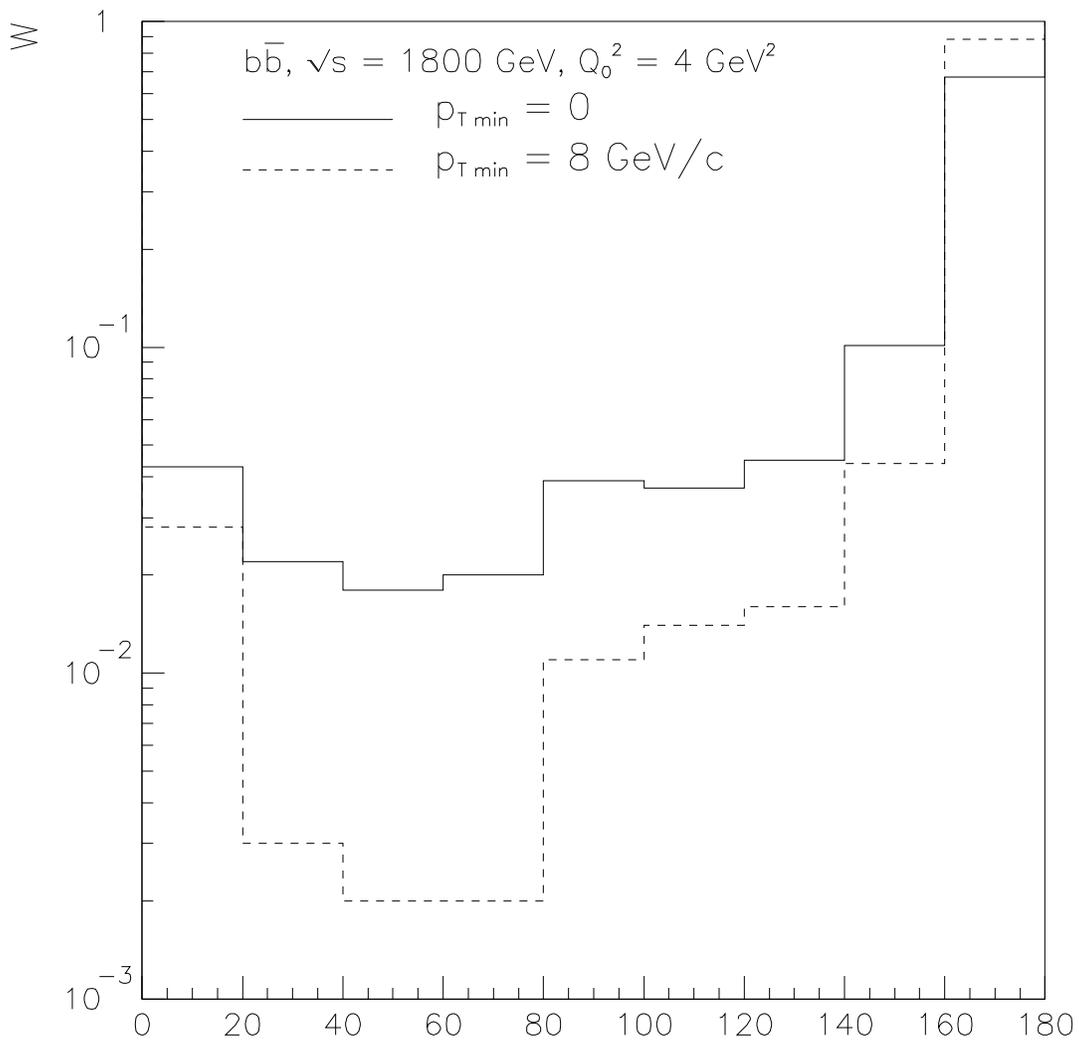
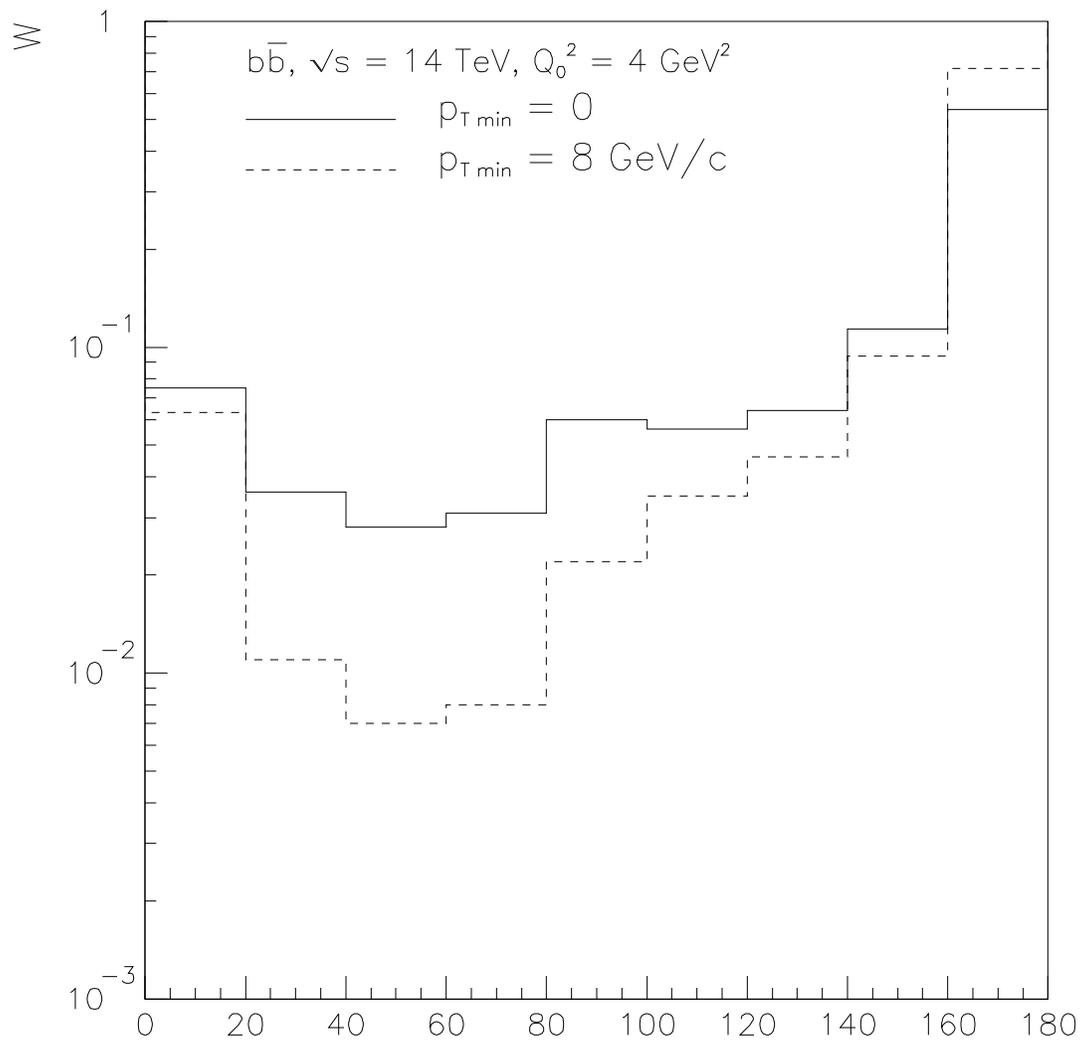
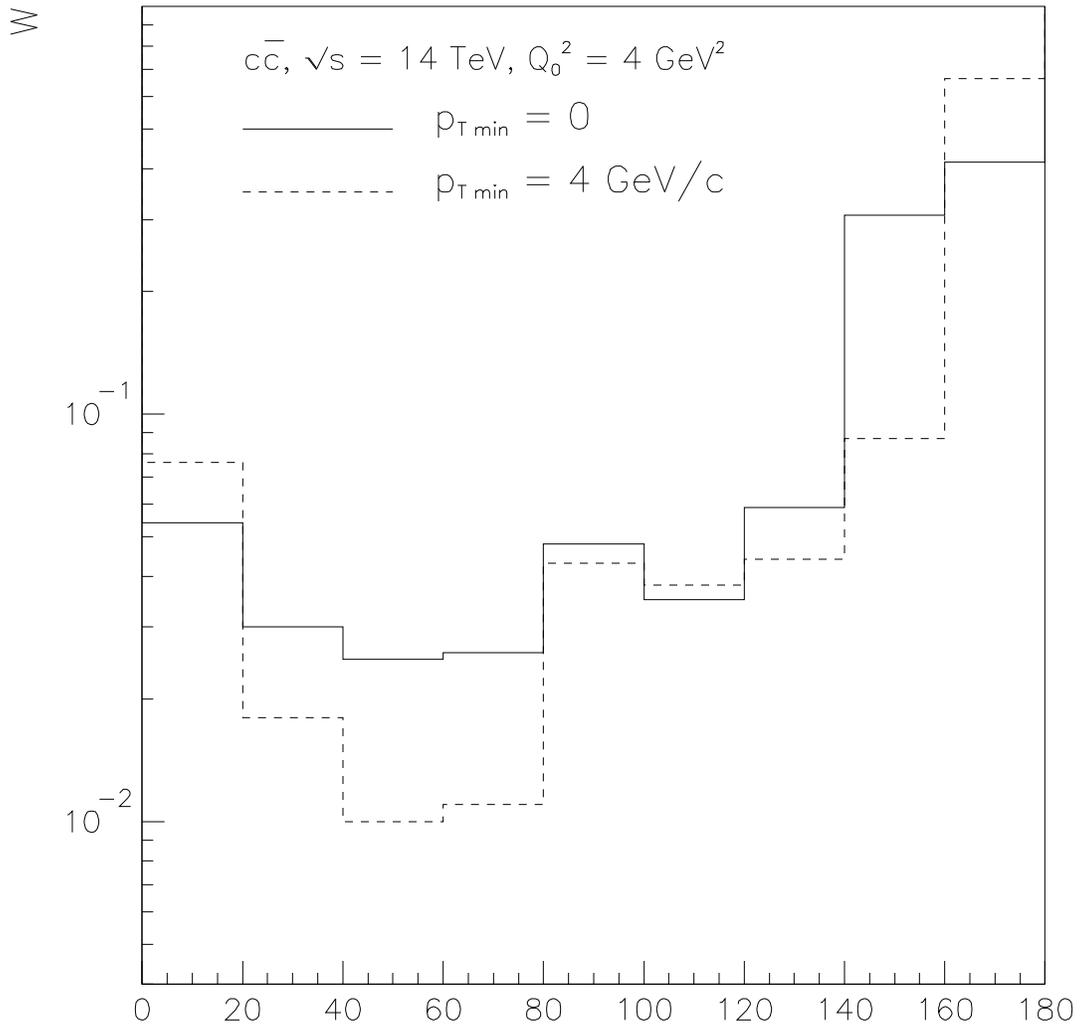
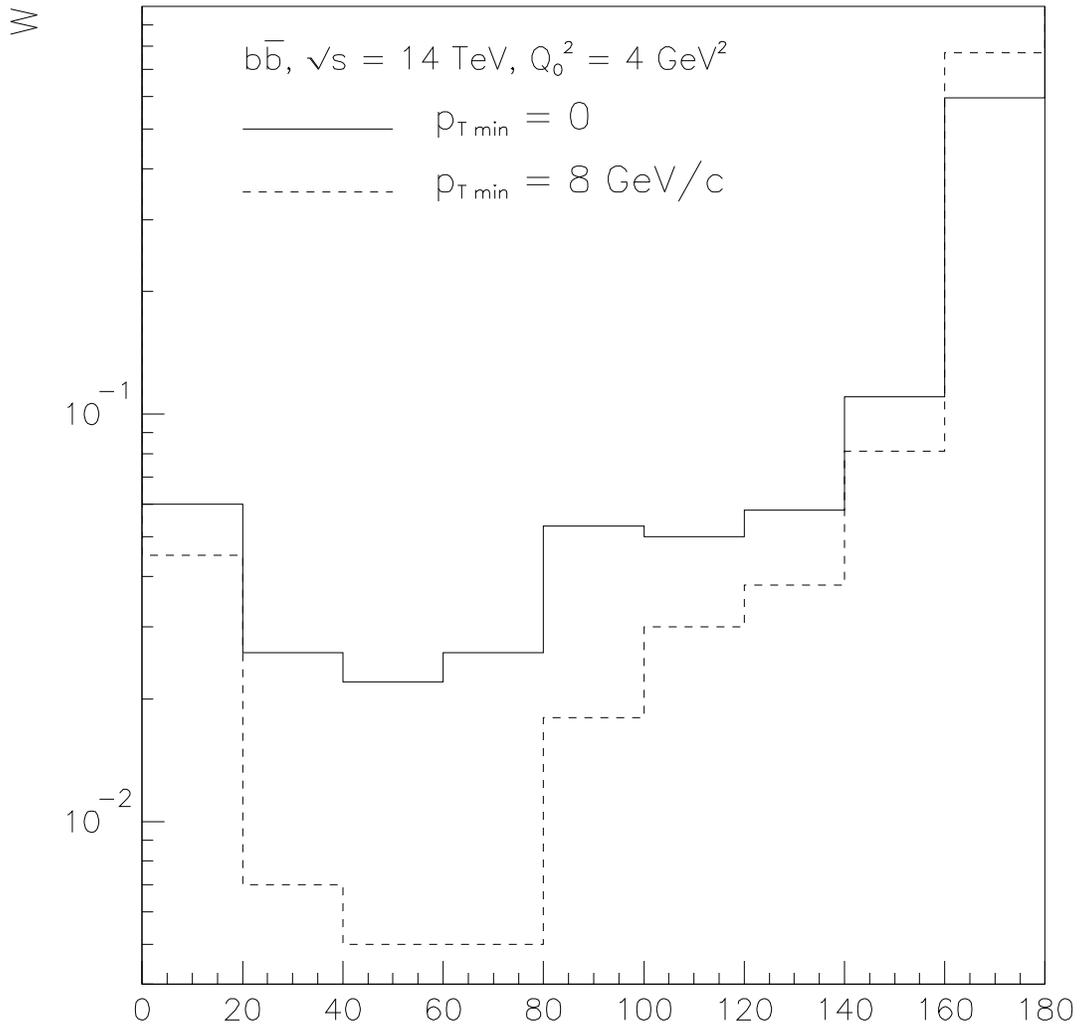
\end{document}